\documentclass[twocolumn,prl,showpacs]{revtex4}
\usepackage{graphicx}
\begin{document}

\title{
Superconductivity in the New Platinum Germanides
$M$Pt$_4$Ge$_{12}$ \\ ($M$ = Rare-earth and Alkaline-earth Metals) \\
with Filled Skutterudite Structure
}

\author{R.\ Gumeniuk}
\author{W.\ Schnelle}
\author{H.\ Rosner}
\author{M.\ Nicklas}
\author{A.\ Leithe-Jasper}
\email{jasper@cpfs.mpg.de}
\author{Yu.\ Grin}

\affiliation{Max-Planck-Institut f\"ur Chemische Physik fester Stoffe,
N\"othnitzer Str.\ 40, 01187 Dresden, Germany}

\begin{abstract}
New germanium-platinum compounds with the filled-skutterudite
crystal structure were synthesized. The structure and composition
were investigated by X-ray diffraction and microprobe analysis.
Magnetic susceptibility, specific heat, and electrical resistivity
measurements evidence superconductivity in LaPt$_4$Ge$_{12}$ and
PrPt$_4$Ge$_{12}$ below 8.3\,K. The parameters of the normal and
superconducting states were established. Strong coupling and a
crystal electric field singlet groundstate is found for the Pr
compound. Electronic structure calculations show a large density of
states at the Fermi level. Similar behavior with lower
$T_\mathrm{c}$ was observed for SrPt$_4$Ge$_{12}$ and
BaPt$_4$Ge$_{12}$.
\end{abstract}

\pacs{74.10.+v, 74.25.Bt, 74.25.Jb \\
submitted to Physical Review Letters July 12, 2007}

\maketitle

Due to the wealth of topical behaviors, compounds with crystal
structures composed of rigid covalently bonded cage-forming
frameworks enclosing differently bonded guest atoms currently
attract much attention. These are in particular the families of
skutterudites and clathrates. The increasing interest and efforts to
understand the underlying physics and chemistry is motivated by the
fascinating diversity of observed physical properties which are due
to subtle host-guest interactions and are, moreover, accessible to
tuning. For the skutterudites they range from magnetic and
quadrupole ordering to heavy-fermion and non-Fermi liquids,
fluctuating valency, superconductivity, itinerant ferromagnetism,
and even half-metallicity
\cite{Uher01,SalesREHandbook,LeitheJasper03etal}. Also, promising
applications as thermoelectric materials considerably boost the
popularity of these compounds \cite{Nolas99,Uher03}.

The generic class of ``skutterudites'' derives from the archetypal
mineral skutterudite CoAs$_3$. The general formula of filled
skutterudites is $M_yT_4X_{12}$, where $M$ is an electropositive
cation, $T$ -- until now -- a transition metal of the iron or cobalt
group, and $X$ a pnicogen (P, As, or Sb). The cations $M$ reside in
icosahedral cages formed by tilted [$TX_6$] octahedra with
inter-octahedral $X$--$X$ bonds. The bonding situation in filled
skutterudites can be understood as an electron donation from the
guest atoms $M$ to the polyanionic framework
\cite{Uher01,LeitheJasper04etal}. The binary skutterudites of cobalt
group elements are diamagnetic semiconductors with a valence
electron count of 72 per [$T_4X_{12}$] formula unit (assuming a
two-center two-electron bonding for $T$--$X$ and $X$--$X$ and a
spin-paired $d^6$ configuration for $T^{3+}$) \cite{Uher01}. In the
iron-group metal-pnicogen skutterudites the $d^5$-configuration
of $T^{3+}$ is not any more sufficient, and cations have to be
embedded. The structure motif becomes stabilized for
NaFe$_4$Sb$_{12}$ already with 70 electrons per f.u.\
\cite{LeitheJasper04etal}. The resulting electron deficiency induces
metallicity associated with paramagnetism and in some cases
collective magnetism \cite{Uher01,SalesREHandbook,LeitheJasper03etal}.

The aforementioned facts suggest a certain flexibility of the
crystal structure with respect to the total electron numbers. There
are several ways of tuning the electron count and thus the physical
properties of skutterudites: \textit{i}) by the selection of an
appropriate guest; \textit{ii}) by replacing part of the pnicogen
atoms by elements of the 14th or 16th group; \textit{iii}) by
substituting the transition-metal atoms, or by combining these
approaches. This chemical access to the electronic properties proved
already extremely successful in the quest to enhance the
thermoelectric properties of skutterudites \cite{Nolas99,Uher03},
and moreover provides a high degree of freedom for explorative
synthetic work. Thus, from this point of view it seemed particularly
interesting if a skutterudite structure could be stabilized composed
of other cage-forming elements than pnicogens. The recent resurgence
in interest in germanium-based clathrates and the promising
electronic applications of these framework structures
\cite{Cohn99,Connetable03,Moriguchi00,Guloy06a} motivated us to
search for chemically compatible compounds which might be accessible
to standard semiconductor technologies. Indeed, there exist several
skutterudite-related structures with cobalt-group transition-metals
where the pnicogen atoms are replaced by one half each of Ge or Sn
and group 16 elements thus retaining the electronic balance
\cite{Nolas03,Bos07}.

In this Letter we report our successful efforts to synthesize the
new skutterudite-like compounds $M$Pt$_4$Ge$_{12}$ composed of
germanium and, for electronic balance reasons, platinum as a
transition metal. They accommodate La and Pr as well as other
elements (e.g.\ $M$ = Sr, Ba, Ce, Nd, Eu) \cite{Gumeniuktobe} as
cations. We examined the basic structural properties
through X-ray powder diffraction (XRD) and electron probe
microanalysis (EPMA). The physical properties were determined
through thermodynamic and transport measurements. We observe the
appearance of superconductivity with transition temperatures
$T_\mathrm{c}$ of 8.3\,K for LaPt$_4$Ge$_{12}$ and -- unexpectedly
high -- 7.9\,K for PrPt$_4$Ge$_{12}$ containing 4$f$ electrons. Our
experimental data are supported by results of full-potential band
structure calculations using the local density approximation. Though
SrPt$_4$Ge$_{12}$ and BaPt$_4$Ge$_{12}$ are also superconducting we
will only describe in detail the two most prominent superconductors.

\begin{table}[ht]
\begin{center}
\caption{Crystallographic data of superconducting cubic $M$Pt$_4$Ge$_{12}$
at room temperature. Structure type: LaFe$_4$P$_{12}$. Space group:
$Im\bar{3}$, $Z$ = 2. Parameters of the superconducting state were
derived from the specific heat data.
\label{thetable}}
\begin{ruledtabular}
\begin{tabular}{lccccc}
$M$ & $a$/\AA   & $T_\mathrm{c}$/K & $\gamma_\mathrm{N}$/mJ\,mol$^{-1}$\,K$^{-2}$ & $\Delta c_p/\gamma_\mathrm{N}T_\mathrm{c}$ & $2\Delta/k_\mathrm{B}T_\mathrm{c}$ \\ \hline
La  & 8.6235(3) & 8.27             & 75.8                                         & 1.49                                       & 1.94                               \\
Pr  & 8.6111(6) & 7.91             & 87.1                                         & 1.56                                       & 2.35                               \\
Sr  & 8.6509(5) & $\approx$5.4     & 39.9                                         & 1.13                                       & --                                 \\
Ba  & 8.6838(5) & $\approx$5.0     & 34.0                                         & 1.17                                       & --                                 \\
\end{tabular}
\end{ruledtabular}
\end{center}
\end{table}
\noindent

Samples were prepared by standard techniques. Elements in
stoichiometric amounts were arc-melted on a water-cooled copper
hearth inside a glove box with Ar-atmosphere ($O_2$ and H$_2$O
$<1$\,ppm). A high-frequency furnace was used for melting the educts
of the Sr and Ba compounds. The compounds were obtained after
annealing at 800\,$^\circ$C for 7\,d in sealed Ta ampoules.
Metallographic and EPMA investigations of polished specimens
revealed elemental Ge and traces of PtGe$_2$ ($<$1\,vol{\%}) as the
only impurity phases in the rather porous samples.

EPMA confirmed the ideal composition
(La$_{0.95(10)}$Pt$_{3.9(1)}$Ge$_{12.2(2)}$) of the compound. No
indications of a homogeneity range La$_y$Pt$_4$Ge$_{12}$ were found:
lattice parameters of samples with $y$ = 0.9, 1.0, 1.1 are equal
within 3 standard deviations. No defect occupation of the cation
position was found by full-profile refinements from powder XRD data
(for lattice parameters see Table \ref{thetable}).

Magnetization was measured down to 1.8\,K in a SQUID magnetometer
(MPMS XL-7, Quantum Design) in various external fields. The heat
capacity was determined by a relaxation-type method on a PPMS
(Quantum Design). In the same measurement system electrical
resistivity data (ac, $\nu$ = 13.4\,Hz) were collected with a
nominal current density of 0.06\,A\,mm$^{-2}$ in several magnetic
fields transverse to the current.

\begin{figure}[htb]
%%%%%%%%%%%%%%
\includegraphics[height=3.4in,angle=90]{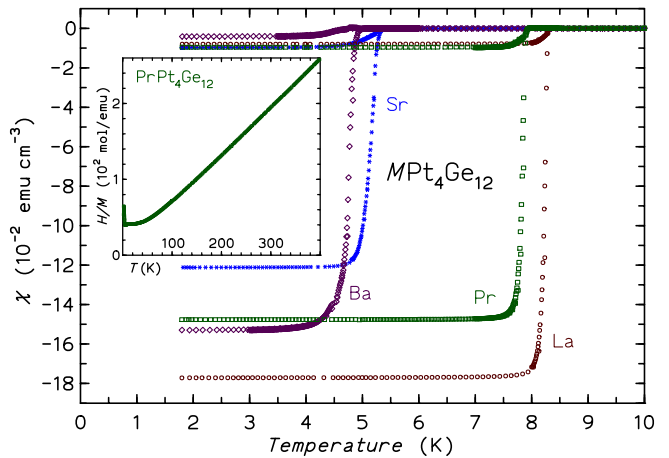}
\caption{(color online) Magnetic volume susceptibility $\chi(T)$ of
filled skutterudites $M$Pt$_4$Ge$_{12}$ ($M$ = Sr, Ba, La, Pr)
for $\mu_0H$ = 2\,mT (nominally). The inset shows the inverse molar
susceptibility $H/M$ of the Pr compound for $\mu_0H$ = 1\,T.
\label{figm}}
\end{figure}

The low-field susceptibility (Fig.\ \ref{figm}) displays strong
diamagnetic signals due to superconducting transitions with onset at
8.29(2)\,K (La), 7.92(2)\,K (Pr), 5.45(8)\,K (Sr), and 4.98(10)\,K
(Ba). While shielding by supercurrents comprises the whole sample
volume (zero-field cooled curves, considering demagnetizing factor
and the high porosity of the samples) the field-cooling Meissner
effect is less than unity. This incomplete Meissner effect is well
known to be due to strong pinning. The inset of Fig.\ \ref{figm}
shows the paramagnetic susceptibility of the Pr compound.
Normal-state $M/H$ ($\mu_0H > H_\mathrm{c2}(1.8$\,K)) is finite at
low temperatures indicating a non-magnetic crystal electric field
(CEF) groundstate for the praseodymium ions. At high temperatures
the effective magnetic moment is 3.55\,$\mu_\mathrm{B}$ (Weiss
temperature $\theta_\mathrm{P} = -$6.7(2)\,K) confirming the
presence of Pr in 4$f^2$ configuration, i.e.\ in the trivalent
state. The isostructural compounds NdPt$_4$Ge$_{12}$ and
EuPt$_4$Ge$_{12}$ with Nd$^{3+}$ and Eu$^{2+}$ ions, respectively
(Kramers ions with magnetic CEF groundstates), are not
superconducting above 0.48\,K and display magnetic order at
0.67\,K and 1.7\,K \cite{Gumeniuktobe}, in spite of stronger
antiferromagnetic interactions (i.e.\ for $M$ = Eu,
$\theta_\mathrm{P}$ = $-$18.4\,K for $\mu_0H$ = 0.01\,T from a
low-$T$ Curie-Weiss fit.

\begin{figure}[htb]
%%%%%%%%%%%%%%
\includegraphics[width=3.4in,angle=0]{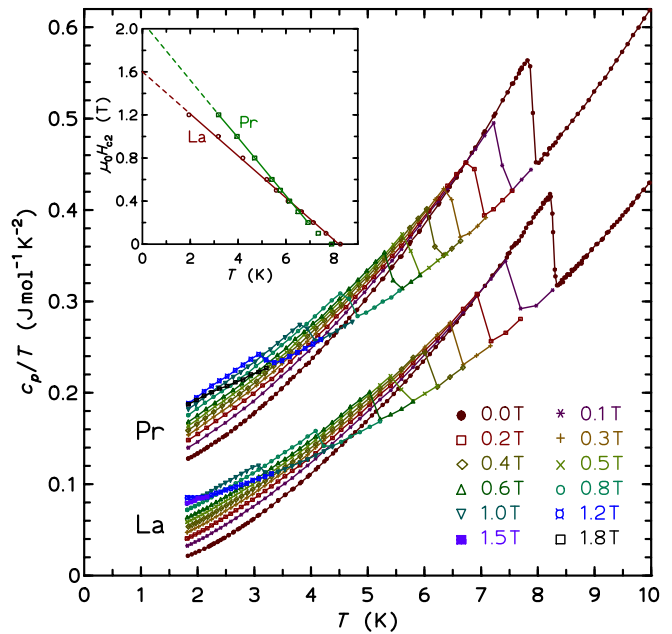}
\caption{(color online) Molar specific heat $c_p/T$ of
LaPt$_4$Ge$_{12}$ and PrPt$_4$Ge$_{12}$ (shifted upwards by 0.1
units) for different magnetic fields. The inset shows the critical
fields $H_\mathrm{c2}$ derived from the midpoints of the transitions
in $c_p(T,H)$.
\label{figcp}}
\end{figure}

The specific heats of the two superconductors with the highest
$T_\mathrm{c}$ (the La and Pr compounds) in various magnetic fields
are presented in Fig.\ \ref{figcp}. The normal state specific heat
of the La compound is well described by a specific model previously
applied to other filled skutterudite compounds
\cite{Keppens98,Schnelle05a,LeitheJasper07a,Schnelletobe}. The
results of these fits suggest that -- similar as in
[Fe$_4$Sb$_{12}$]-based skutterudites -- the La atom's vibrations
are well described by a low-energy Einstein term while the
polyanionic host [Pt$_4$Ge$_{12}$] can be treated by the
low-temperature Debye approximation $c \propto T^3$. The fit in the
range of 3--10\,K results in a Debye temperature $\Theta_\mathrm{D}$
= 209\,K for the polyanion (16 atoms), Einstein temperature
$\Theta_\mathrm{E}$ = 96\,K for the cation, and the Sommerfeld
coefficient $\gamma_\mathrm{N}$ = 76\,mJ\,mol$^{-1}$\,K$^{-2}$. For
the Pr compound the values are $\Theta_\mathrm{D}$ = 198\,K and
$\gamma_\mathrm{N}$ = 87\,mJ\,mol$^{-1}$\,K$^{-2}$
($\Theta_\mathrm{E}$ cannot be determined unambiguously due to the
CEF contribution of Pr). $\Theta_\mathrm{D}$ = 198\,K for Sr and
209\,K for BaPt$_4$Ge$_{12}$ are very similar, as it can be expected
for the same polyanion.

The jumps in $c_p/T$ at $T_\mathrm{c}$ were determined by the usual
entropy-conserving construction. The relative changes of $\Delta
c_p/\gamma_\mathrm{N}T_\mathrm{c}$ are 1.49 and 1.56 for La and
PrPt$_4$Ge$_{12}$, respectively. While this ratio indicates an
coupling somewhat stronger than in the BCS theory ($\Delta
c_p/\gamma_\mathrm{N}T_\mathrm{c}$ = 1.426) for LaPt$_4$Ge$_{12}$,
the Pr compound is clearly a strong coupling superconductor.
Comparing the electronic specific heat $c_\mathrm{eS}(T)$ for $H =
0$ at $T_\mathrm{c}/T$ = 2 with the $\alpha$-model \cite{Padamsee73}
we obtain energy gap ratios $2\Delta/k_\mathrm{B}T_\mathrm{c}$ of
1.94 (La) and 2.35 (Pr) (BCS theory
$2\Delta/k_\mathrm{B}T_\mathrm{c}$= 1.764), confirming the strong
coupling in the Pr compound.

\begin{figure}[htb]
%%%%%%%%%%%%%%
\includegraphics[height=3.4in,angle=90]{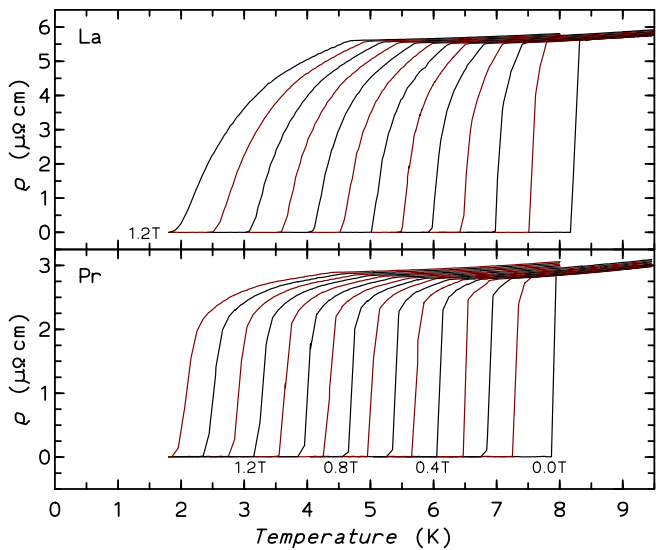}
\caption{(color online)
Electrical resistivity of LaPt$_4$Ge$_{12}$ (La) and
PrPt$_4$Ge$_{12}$ (Pr) in magnetic fields in steps
$\Delta\mu_0H$ of 0.1\,T.
\label{figrho}}
\end{figure}

The field dependence of the upper critical field $H_\mathrm{c2}$ was
determined from the midpoints of the jumps in $c_p(T,H)$. The
results are given in the inset of Fig.\ \ref{figcp} and suggest that
$\mu_0H_\mathrm{c2}$ varies almost linearly with $T$. In order to
confirm this finding the electrical resistivity was measured in
several magnetic fields. Results are depicted in Fig.\ \ref{figrho}
for equally spaced field intervals of $\Delta\mu_0H$ = 0.1\,T. It is
obvious from the zero-resistance points of the curves ($\rho(T,H) =
0$) that $T_\mathrm{c}(H)$ varies linearly with the applied field.
An extrapolation (see Fig.\ \ref{figcp} inset) yields
$\mu_0H_\mathrm{c2}(0)$ = 1.60\,T and 2.06\,T for La and
PrPt$_4$Ge$_{12}$, respectively.

The residual and room temperature resistivities of the current
porous polycrystalline samples are surprisingly low: 5.3 and 177
$\mu\Omega$\,cm for $M$ = La and 2.6 and 110 $\mu\Omega$\,cm for $M$
= Pr. Therefore the superconductivity is assumed to be in the clean
limit. The magnetoresistance ratio $(\rho(H)-\rho(0))/\rho(0)$
increases strongly with decreasing $T$. For the La compound it
amounts to +66\,{\%} at $\mu_0H$ = 9\,T (+23\,{\%} at 4\,T; +9\,{\%}
at 2\,T) just above $T_\mathrm{c}$. PrPt$_4$Ge$_{12}$ displays
similar values (+83\,{\%} in 9\,T).

The presence of superconductivity in a compound containing
paramagnetic rare-earth ions is usually destroyed by pair breaking.
The CEF leading to a non-magnetic singlet groundstate in
PrPt$_4$Ge$_{12}$ can be estimated by calculating the excess
specific heat $c_\mathrm{CEF}(T)$ with respect to the La compound.
The resulting heat capacity can be fitted with a Schottky anomaly
with an energy level scheme for the $^3H_4$ ground multiplet of the
4$f^2$ state of Pr$^{3+}$ on a site with cubic point symmetry $T_h
(m3)$ \cite{Takegahara01}. We find a $\Gamma_1$ singlet as
groundstate, the non-magnetic doublet $\Gamma_{23}$ at a splitting
of $\Delta E/k_\mathrm{B}$ = 93(5)\,K and the two triplets
$\Gamma_4$ at 159(10)\,K and 170(20)\,K, respectively. The
well-isolated non-magnetic CEF groundstate of the Pr obviously
leaves the superconductivity in PrPt$_4$Ge$_{12}$ almost
undisturbed. For the isostructural compound PrRu$_4$Sb$_{12}$ with a
CEF splitting $\Delta E/k_\mathrm{B}$ = 125\,K the $T_\mathrm{c}$ is
well below that of the corresponding La compound ($T_\mathrm{c}$ =
1.04\,K \textit{vs.} 3.58\,K) \cite{Takeda00}. Other La and
Pr-filled superconducting skutterudite phases based on iron group
metals with pnicogen atoms were also reported. While for
La$_x$Rh$_4$P$_{12}$ $T_\mathrm{c}$ up to 17\,K is found the
corresponding Pr compound has only a $T_\mathrm{c}$ of 2.4\,K
\cite{Shirotani05}. Thus, PrPt$_4$Ge$_{12}$ is a ``van-Vleck
superconductor'' with a remarkably high $T_\mathrm{c}$. In this
respect, the reasons for the low magnetic ordering temperatures in
the Nd and Eu compounds merit further investigation.

\begin{figure}[htb]
%%%%%%%%%%%%%%
\includegraphics[height=3.4in,angle=270]{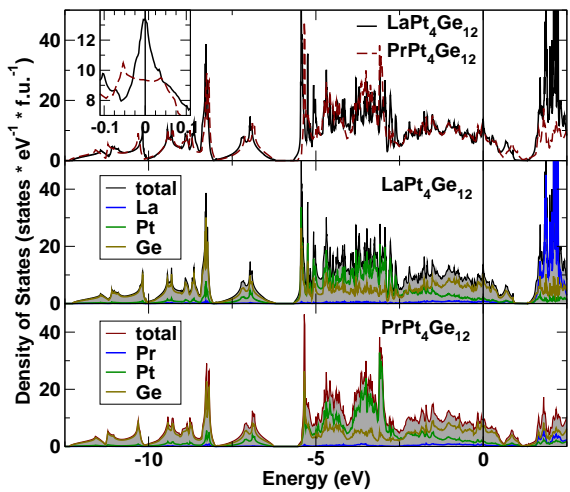}
\caption{(color online)
Total (upper panel) and atom resolved electronic density of states for
LaPt$_4$Ge$_{12}$ and PrPt$_4$Ge$_{12}$. The inset in the upper panel
shows the well-pronounced peak in LaPt$_4$Ge$_{12}$ in a narrow region
around the Fermi level.
\label{figdos}}
\end{figure}

To study the electronic structure of La(Pr)Pt$_4$Ge$_{12}$ a
full-potential non-orthogonal local-orbital calculation scheme
\cite{FPLOKoepernik99} within the local density approximation
was applied. In the full-relativistic calculations, the exchange and
correlation potential of Perdew and Wang \cite{PerdewWang92} was
used. As the basis set, La(5$s$, 5$p$, 6$s$, 6$p$, 5$d$, 4$f$), Pr
(5$s$, 5$p$, 6$s$, 6$p$, 5$d$), Pt(5$s$, 5$p$, 6$s$, 6$p$, 5$d$),
and Ge (3$d$, 4$s$, 4$p$,4$d$) states were employed. All lower-lying
states were treated as core states \cite{RemarkPr}. A very dense
$k$-mesh of 3333 points in the irreducible part of the Brillouin
zone (74088 in the full zone) was used to ensure accurate density of
states information, especially in the region close to the Fermi
level. An optimization of the cubic crystal structure ($a$, $y$(Ge),
$z$(Ge)) for the La compound resulted in very good agreement with
the powder refinement emphasizing the structural stability of this
system.

The calculated total density of states (DOS) for LaPt$_4$Ge$_{12}$
in comparison to PrPt$_4$Ge$_{12}$ is plotted in Fig.\ \ref{figdos}
(upper panel). On first glance, the electronic DOS of both compounds
is very similar. The contribution of La and Pr to the valence band
are very small and featureless. The low-lying bands between about
$-$12 eV and $-$6\,eV are predominantly bonding Ge 6$s$ states. The
majority of Pt 5$d$ states is located between about $-$5.5\,eV and
$-$2.5\,eV and hybridizes strongly with the Ge $4p$ orbitals. These
facts strongly support the picture of a charge transfer from the
rare-earth cation to the [Pt$_4$Ge$_{12}$] polyanion. The states at
the Fermi level $\varepsilon_\mathrm{F}$ are mainly (about 80\,{\%})
due to Ge 4$p$ bands. At $\varepsilon_\mathrm{F}$ the DOS values are
$\gamma_0$ = 13.4 and 9.3 states eV$^{-1}$ f.u.$^{-1}$ for the La
and the Pr compound, respectively, corresponding to coupling
constants $\lambda$ = $\gamma_\mathrm{N}/\gamma_0 - 1$ of 1.4 and
2.9. This gives further evidence for the considerably stronger
coupling in the PrPt$_4$Ge$_{12}$.

In conclusion, we have found new intermetallic compounds of platinum
and germanium with the filled-skutterudite crystal structure. The
La, Sr, Ba and -- most noteworthy -- the Pr-filled compounds are
superconductors with maximum $T_\mathrm{c}$ around 8\,K.
Thermodynamic measurements indicate strong coupling
for PrPt$_4$Ge$_{12}$ and a non-standard variation of
$H_\mathrm{c2}$ with temperature.

\begin{acknowledgments}
We are indebted to U.\ Burkhardt, T.\ Vogel, and R.\ Koban for
assistance and S.-L.\ Drechsler and L.\ Akselrud for valuable
discussions. H.\ R.\ acknowledges the DFG (Emmy-Noether-Programm)
for financial support.
\end{acknowledgments}


\begin{thebibliography}{24}
\bibitem[{Rem()}]{RemarkPr}
\bibinfo{note}{{The fully localized treatment of the Pr 4$f$ is well justified
  due to the evident Pr$^{3+}$ state from the experimental data.}}
\end{thebibliography}
\end{document}